\documentstyle[twoside,fleqn,espcrc2]{article}

\newcommand{\ttbs}{\char'134}
\newcommand{\AmS}{{\protect\the\textfont2
  A\kern-.1667em\lower.5ex\hbox{M}\kern-.125emS}}

\title{Performance of the Cray T3D and Emerging Architectures
	on Canopy QCD Applications}

\author{Mark Fischler and Mike Uchima
	\address{Computing Division, Fermilab,
        Batavia, IL 60510-0500, U.S.A}%
        \thanks{Fermilab is operated by Universities Research Association, Inc.
		under contract with the U.S. Department of Energy.}
	} 

\begin{document}

\begin{abstract}
The Cray T3D, an MIMD system with NUMA shared memory capabilities and in
principle very low communications latency, can support the Canopy framework
for grid-oriented applications.
CANOPY has been ported to the T3D,
with the intent of making it available to a spectrum of users.
The performance of the T3D running Canopy has been benchmarked on five QCD
applications extensively run on ACPMAPS at Fermilab, requiring a variety of
data access patterns.
The net performance and scaling behavior
reveals an efficiency relative to peak Gflops almost identical to that
achieved on ACPMAPS.

Detailed studies of the major factors impacting performance are presented.
Generalizations applying this analysis to the newly emerging crop
of commercial systems reveal where their limitations will lie.
On these applications, efficiencies of above 25\% are not to be expected;
eliminating overheads due to Canopy will improve matters, but by less than
a factor of two.
\end{abstract}

\maketitle

\section{T3D Canopy}

The Cray T3D is a massively parallel MIMD system with processing elements (PEs)
containing 150 Mflop DEC Alpha CPUs.
It has distributed memory, with processing units (PEs) connected in a
3-D periodic grid.
Access to memories associated with remote PEs is provided via shared
memory calls (``shmem'' routines \cite{shmem});
this allows for the flat global access
paradigm assumed by the underpinnings of the Canopy \cite{Canmanual} framework
for grid-oriented applications.
Moreover, this access has low latency:
3 $\mu$sec round trip (to specify data required and get back the data)
in principle; 7 $\mu$sec for a read access in a repeated ``ping-pong'' test;
and 23$\pm$13 $\mu$sec net cost per transfer during actual applications.
The MIMD architecture, with low-latency flat global access,
makes the T3D well suited as a Canopy platform.
In cooperation with Cray Research, Inc. and the Pittsburgh Supercomputing
Center, we have ported Canopy to this architecture,
and tested performance on the 512-PE system at PSC.

The porting strategy was straightforward:
Implement remote access routines in terms of the
shared memory primitives (e.g. {\tt shmem$\_$get}) to
create the CHIP-level (Canopy Hardware Interface Package),
and clean up any Canopy dependence
on 32-bit addresses.
The higher-level routines in the Canopy
library are written in terms of these underpinnings,
and required little additional effort once
CHIP was implemented.

Some issues involving implementing remote access in terms of the
shmem routines:
The paradigm assumes memory and cache coherency and access ``causality'';
this behavior can be selected by specifying automatic
invalidation of cache lines under remotely written data, and
by using the {\tt write$\_$barrier} mechanism in implementing
{\tt remote$\_$write}.

\begin{table*}[hbt]
\setlength{\tabcolsep}{1.5pc}
\newlength{\digitwidth} \settowidth{\digitwidth}{\rm 0}
\catcode`?=\active \def?{\kern\digitwidth}
\caption{Per-PE performance (Mflops) of T3D and ACPMAPS on QCD benchmarks}
\label{tab:effs}
\begin{tabular*}{\textwidth}{@{}l@{\extracolsep{\fill}}crrcrr}
\hline
	& \multicolumn{3}{c}{Single PE}
	& \multicolumn{3}{c}{128 PEs} \\
\cline{2-4} \cline{5-7}
                 & \multicolumn{1}{c}{\underline{Cray T3D}}
                 & \multicolumn{2}{c}{ACPMAPS}
                 & \multicolumn{1}{c}{\underline{Cray T3D}}
                 & \multicolumn{2}{c}{ACPMAPS} \\
\cline{3-4} \cline{6-7}
1---Pure Gauge		& $ 7.7$ & $5.1$ & $9.8$ & $ 7.2$ & $4.3$ & $7.1$ \\
2---Conj. Grad.		& $ 7.9$ & $4.1$ &$14.0$ & $ 6.1$ & $3.4$ & $8.5$ \\
3---MinRes L--U		& $ 7.6$ & $4.2$ &$12.9$ & $ 2.9$ & $1.9$ & $3.4$ \\
4---Gauge Fix		& $11.0$ & $6.2$ &$12.2$ & $ 9.3$ & $5.1$ & $8.6$ \\
5---FFT Gauge Fix	& $11.4$ & $6.1$ & $9.0$ & $ 7.1$ & $3.5$ & $4.4$ \\
\hline
\multicolumn{7}{@{}p{160mm}}
{Benchmarks are described in the text. Second value for ACPMAPS is with
optimized kernels.}
\end{tabular*}
\end{table*}


The T3D's strengths as a Canopy platform for QCD include:
The proper memory model (flat global
access) was easily implemented in terms of vendor-supplied options;
the Alpha CPU has good performance over a wide range of activities;
and the communications features excellent interprocesor
bandwidth, good per-transfer overhead, and negligible degradation when
transfers
are non-nearest-neighbor.
Performance scales well, up to very large numbers of PEs.

The major weakness is the size of data and instruction caches, which are
8K each on chip, with no second-level cache.
QCD inevitably incurs frequent data cache misses,
but this instruction cache is too small to avoid costly instruction misses.
When the T3D is used as a substitute for a ``long vector'' machine---as many
Cray customers will do---the instruction cache is adequate for the
dominant simple loops.
But for Canopy implementations of our QCD
benchmarks, the impact of instruction thrashing is significant,
to the extent that performance values can
fluctuate by a few percent depending on the order in which routines were
linked.

The other weakness is subroutine calling overhead:  The T3D supports
traceback capabilities, at the cost of quite a few extra instructions per call.
This penalizes the modular organization emphasized by Canopy.
Cray is preparing options to avoid much of this overhead where
appropriate.

\section{Performance Benchmarks}

Benchmarking on ``clean'' codes is risky---simple
applications can lead to deceptive results.
Instead, it is best to use a suite of actual production codes.
We have selected a suite consisting of the five most heavily used production
QCD
codes run on ACPMAPS.
Each of these applications has been run more than
1000 sustained Gflop-hours at
Fermilab.
Fortuitously, these also display a variety of characteristics in terms of
communication and computation.

The applications are:
(1) Kennedy-Pendleton heat-bath gauge configuration generation---fairly local
with moderate communication burden;
(2) DeGrand Conjugate Gradient propagator computation---higher
communication frequency;
(3) MRLU Minimum Residual Incomplete LU-preconditioned propagator
computation---still more communication, with
complicated access and synchronization patterns;
(4) Relaxation method Coulomb gauge fixing;
(5) FFT-accelerated Coulomb gauge fixing---significant non-local
data traffic.
The benchmark results are shown in table \ref{tab:effs}.

\begin{table*}[bt]
\setlength{\tabcolsep}{1.5pc}
\settowidth{\digitwidth}{\rm 0}
\catcode`?=\active \def?{\kern\digitwidth}
\caption{Sources of inefficiency in a QCD calculation}
\label{tab:effcosts}
\begin{tabular*}{\textwidth}{@{}l@{\extracolsep{\fill}}rrrr}
\hline
                 & \multicolumn{2}{c}{Cray T3D}
                 & \multicolumn{2}{c}{Emerging Systems} \\
                 & \multicolumn{2}{c}{({\em ``predicted''})}
                 & \multicolumn{2}{c}{{\em median (worst)}} \\
\cline{2-3} \cline{4-5}
                 & \multicolumn{1}{r}{Inherent}
                 & \multicolumn{1}{r}{+ Canopy}
                 & \multicolumn{1}{r}{Inherent}
                 & \multicolumn{1}{r}{+ Canopy} 		\\
Lack of kernel optimization       & $245$ &  ---  & 	 		\\
Unavoidable kernel inefficiency   & $ 60$ & $ 35$ & $ 70$ & $ 40$ 	\\
C-compiled non-kernel flops       & $ 45$ &  ---  & $ 15$ & ---		\\
Subroutine overheads		  &  ---  & $ 60$ & ---   & $ 10$	\\
Bookkeeping (data-finding, \dots) & $  5$ & $ 80$ & $ 10$ & $ 85$	\\
Loops and result integration      & $  5$ & $ 90$ & $  5$ & $ 25$	\\
Local (main) memory bandwidth	  & $130$ &  ---  & $ 50 (275)$ & --- 	     \\
Local (main) memory latency    	  & $ 15$ & $ 20$ & $ 15 (40)$ & $ 15 (45)$  \\
Communication latency/overhead	  & $ 45$ & $150$ & $  5 (75)$ & $ 20 (225)$ \\
Communication bandwidth		  & $  5$ &  ---  & $ 25 (75)$ & ---	     \\
I-cache misses, cache BW, \dots   & $ 30$ & $200$ & $  5 (25)$ & ---	     \\
\cline{1-1} \cline{2-2} \cline{3-3} \cline{4-4} \cline{5-5}
Expected Efficiency (\% of peak)  & $14.5\%$ & $7.5\%$ &
		$27\pm5\%$ & $18\pm5\%$ 	\\
\hline
\multicolumn{5}{@{}p{120mm}}
{Numbers are normalized to $ideal flops/peak speed = 100\%$.}
\end{tabular*}
\end{table*}

Comparisons to ACPMAPS follow these ground rules:
Key Canopy routines (e.g. {\tt field\-$\_$pointer}) were optimized
on both systems;
application-dependent computational kernels were left in C on
both;
and transfer coalescing \cite{tranco} is used on ACPMAPS,
but not on the T3D where it would be a net loss.
A relevant comparison is between cost per transfer on the T3D
(23$\pm13$ $\mu$sec = 1725 cycles) and the overhead per (coalesced) block
on ACPMAPS (44$\pm14$ $\mu$sec = 1760 cycles).
Note the 50\% fluctuations, which occur for different applications on each
system.

Averaged over applications, the T3D
delivers 1.78 times the (per processor) power of ACPMAPS in single node
performance, and 1.80 times on 128-node jobs.  Since the ratio of peak
power is 150:80 = 1.87, the efficiencies are identical to within 5\%.
The averaged scaling behaviors are also nearly identical, on
1, 2, 4, \dots 128 processors; this is understandable
given the match in cost per transfer.
The T3D scaling behavior is
superior on the FFT (where there are many-transfers-in-sequence steps) and
inferior on MRLU (where multithread transfer coalescing reduces some nodes'
idle time).

Given the similarities in floating point architecture and transfer costs,
we can expect the same performance
improvement on the T3D when computational kernels (e.g. SU(3) multiplication)
are hand-optimized, as were observed on ACPMAPS---an overall factor of 1.8 at
average.  This works out to efficiencies of about 15\%.


The characteristics of the T3D can be put into a general analysis of
expected performance, to test its accuracy.
The ``predicted'' absolute performance and scaling behavior
match the benchmarked behavior at the 20\% level.
Later we will present performance analyses based on characteristics of
several commercial systems
which will emerge in 1996-1998; we are confident at that level in our
estimates of QCD efficiency and of the impact of Canopy.

Table \ref{tab:effcosts} lists the major contributions to cycles
taken (per site) to execute the Wilson fermion CG method.
This application was selected as being similar
to the bulk of work done in full-QCD computations.
Each site, including $\Delta\!\!\!\!/$ operations
(using properties of $\gamma_i$ to cut the multiplies by half) and
dot-product/linear-combination steps, involves 3164 flops.

The major effects which are present irrespective of the Canopy framework are
non-optimal floating-point kernels and memory bandwidth and latency, each
costing roughly 200\%.
The former is largely eliminated when a program is hand optimized; the latter
remains.

Major effects of the Canopy framework
include an increased number of data transfers (though {\em not} an increase
in total traffic), and more ``bookkeeping'' activity.
(Except for methods with intricate synchronization/communication patterns,
the performance cost of Canopy can by assessed by comparing to the time
the same algorithm would take, if one dimension were ``collapsed''.
That is, instead of $N_t$ sites in the time direction, one super-site is used,
so that vectors are long and bookkeeping and transfer overheads are slashed.)
Peculiar to the T3D is the effect of instruction cache misses:
Our applications
are large enough that thrashing occurs; absent Canopy this would be
mitigated by longer computation loops.
The overall cost of Canopy on the T3D is just under a factor of 2.

The measured speed was 20\% lower than these estimates, partly because the C
versions of the kernels
had been distorted to improve performance on the i860, not the Alpha.
Even
removing Canopy costs, and hand-optimizing the
kernels, we would achieve no better than 23\% efficiency for this algorithm
on the T3D.
(The additional heavy computation involved when the ``clover''-improved
$\Delta\!\!\!\!/$ is used might easily push efficiencies into the 30\% range.)

\section{Emerging Systems}

The T3D, with MIMD and flexible, flat global communications capability, is a
precursor to the next wave of commercial MPP systems.
Although specifics are cloaked in non-disclosure secrecy, general trends can be
observed.
The universal leaning (at least among American companies) is toward
massive CPU chip production, so the {\em same chip} that is in
workstations and even PCs will be in the MPP system.
NUMA (Non-Uniform Memory Access) distributed memory systems are becoming
the norm.
System designs are {\em not} being driven by lattice gauge or similar needs;
nonetheless they are good (but not ideal) for QCD, with decent floating point
architectures and high interprocessor bandwidth.
The pleasant surprise is that they are good Canopy platforms, supporting the
remote access paradigm, with low interprocessor latency in most cases.

Positive developments include:  These CPUs have support for shared memory and
cache coherency;  the popular architecture is that of 2 fmac pipes, which does
complex arithmetic well; some systems support prefetch;
and prices will drop driven by the high-volume CPU costs.

On the down side:  Since most mainstream applications rarely miss data cache,
main memory latencies will be dissapointing for QCD, which often misses;
super-features for QCD such as many fmac pipes or 256 registers are not
coming;
and low latency communication is not always a priority.
The biggest headache is that companies will continue to sell their biggest
systems at premium prices, so the cost per flop will remain high.

Of course these systems do message passing, but
they also support remote access communications.
Two such paradigms are supported (each vendor uses different names
for these concepts):  Explicit Remote Access (as assumed in CHIP),
and ``Global Shared Memory'' (GSM) in which an address is
asserted as an ordinary memory load, and the access proceeds transparently.
The latter requires some design cleverness but clearly at least one
company will succeed in delivering it.

GSM can be used to implement Remote Access, and it
opens possibilities of different programming approaches,
and tends toward lower access latencies.
These advantages make GSM superior, but there are some disadvantages:
Because the unit of transfer is a memory cache line
(rather than a specified number of words) some inter\-node bandwidth is wasted
(about 37\% for 64-byte cache lines);
and where GSM requires nodes to share resources,
contention can cost 25--37\% in effective bandwidth.
But the biggest potential drawback
is that the cache checking involved in GSM can seriously
impact access latency to {\em local} memory.

Factors affecting expected performance are addressed in the second half of
table \ref{tab:effcosts}.
There are inefficiencies not dependent on architecture specifics,
amounting to 100\%,
plus other losses ascribable to Canopy amounting to 160\%.
Then for each architecture there are losses from local memory and remote access
limitations.
Median values for each impact are small; but every system has at least one
weak spot, such that the total of these costs ranges from 115--490\%.
Folding this in, we can expect no better than 13--24\% efficiency on Canopy
applications.  And eliminating the Canopy paradigm will only gain a factor of
1.5.

These efficiencies are twice as good as those seen on the T3D;
we have identified where the additional time is lost (to within 20\%).
The relative cost of Canopy will be less on future systems than on the T3D,
because remote access overhead is much smaller and memory bandwidth
issues (independent of Canopy) are more serious.

\section{Summary}

The T3D is a successful Canopy platform, with efficiency and scaling behavior
matching that of ACPMAPS, which was specifically designed for the purpose.
Emerging MPP systems will be good Canopy platforms and decent QCD machines.
But we can't expect more than 25\% efficiency on the best algorithms, or
perhaps 40\% without Canopy.


The expected efficiency and price per peak flop {\em do not\/} correlate
well---the more efficient system is not always the more expensive---so
choosing the correct system will be important.
And although prices will be dropping, getting substantial power will
remain painful---machine design efforts are not yet meritless or obsolete!

\end{document}